\PassOptionsToPackage{unicode}{hyperref}
\PassOptionsToPackage{hyphens}{url}
\PassOptionsToPackage{dvipsnames,svgnames,x11names}{xcolor}
\documentclass[
  letterpaper,
  DIV=11,
  numbers=noendperiod]{scrartcl}
\usepackage{xcolor}
\usepackage[margin=1in]{geometry}
\usepackage{amsmath,amssymb}
\usepackage{iftex}
\ifPDFTeX
  \usepackage[T1]{fontenc}
  \usepackage[utf8]{inputenc}
  \usepackage{textcomp} 
\else 
  \usepackage{unicode-math} 
  \defaultfontfeatures{Scale=MatchLowercase}
  \defaultfontfeatures[\rmfamily]{Ligatures=TeX,Scale=1}
\fi
\usepackage{lmodern}
\ifPDFTeX\else
\fi
\IfFileExists{upquote.sty}{\usepackage{upquote}}{}
\IfFileExists{microtype.sty}{
  \usepackage[]{microtype}
  \UseMicrotypeSet[protrusion]{basicmath} 
}{}
\makeatletter
\@ifundefined{KOMAClassName}{
  \IfFileExists{parskip.sty}{%
    \usepackage{parskip}
  }{
    \setlength{\parindent}{0pt}
    \setlength{\parskip}{6pt plus 2pt minus 1pt}}
}{
  \KOMAoptions{parskip=half}}
\makeatother
\makeatletter
\ifx\paragraph\undefined\else
  \let\oldparagraph\paragraph
  \renewcommand{\paragraph}{
    \@ifstar
      \xxxParagraphStar
      \xxxParagraphNoStar
  }
  \newcommand{\xxxParagraphStar}[1]{\oldparagraph*{#1}\mbox{}}
  \newcommand{\xxxParagraphNoStar}[1]{\oldparagraph{#1}\mbox{}}
\fi
\ifx\subparagraph\undefined\else
  \let\oldsubparagraph\subparagraph
  \renewcommand{\subparagraph}{
    \@ifstar
      \xxxSubParagraphStar
      \xxxSubParagraphNoStar
  }
  \newcommand{\xxxSubParagraphStar}[1]{\oldsubparagraph*{#1}\mbox{}}
  \newcommand{\xxxSubParagraphNoStar}[1]{\oldsubparagraph{#1}\mbox{}}
\fi
\makeatother

\usepackage{longtable,booktabs,array}
\usepackage{calc} 
\usepackage{etoolbox}
\makeatletter
\patchcmd\longtable{\par}{\if@noskipsec\mbox{}\fi\par}{}{}
\makeatother
\IfFileExists{footnotehyper.sty}{\usepackage{footnotehyper}}{\usepackage{footnote}}
\makesavenoteenv{longtable}
\usepackage{graphicx}
\usepackage{pdfpages}
\makeatletter
\newsavebox\pandoc@box
\newcommand*\pandocbounded[1]{
  \sbox\pandoc@box{#1}%
  \Gscale@div\@tempa{\textheight}{\dimexpr\ht\pandoc@box+\dp\pandoc@box\relax}%
  \Gscale@div\@tempb{\linewidth}{\wd\pandoc@box}%
  \ifdim\@tempb\p@<\@tempa\p@\let\@tempa\@tempb\fi
  \ifdim\@tempa\p@<\p@\scalebox{\@tempa}{\usebox\pandoc@box}%
  \else\usebox{\pandoc@box}%
  \fi%
}
\def\fps@figure{htbp}
\makeatother

\providecommand{\tightlist}{%
  \setlength{\itemsep}{0pt}\setlength{\parskip}{0pt}}

\usepackage{fancyhdr}
\usepackage{xcolor}
\usepackage{tikz}

\makeatletter
\@ifundefined{KOMAClassName}{}{%
  \addtokomafont{disposition}{\rmfamily}%
}
\makeatother
\usepackage[explicit]{titlesec}

\definecolor{colleaguepurple}{HTML}{7C6CEB}
\definecolor{keyfindingsbg}{RGB}{254,248,232}

\usepackage[breakable]{tcolorbox}
\newtcolorbox{keyfindings}{
  colback=keyfindingsbg,
  colframe=keyfindingsbg,
  boxrule=0pt,
  arc=16pt,
  left=20pt,
  right=20pt,
  top=12pt,
  bottom=12pt,
  grow to left by=12pt,
  grow to right by=12pt,
  breakable
}

\definecolor{sectionbg}{HTML}{EDF1F7}

\fancypagestyle{plain}{
  \fancyhf{}
  \fancyfoot[L]{\small\nouppercase{\leftmark}}
  \fancyfoot[R]{\small\thepage}

}

\titleformat{\section}
  {\clearpage\normalfont}
  {}
  {0pt}
  {%
    \noindent\begin{tikzpicture}[remember picture, overlay]
      \fill[sectionbg]
        (current page.north west) rectangle
        ([yshift=-3.5cm]current page.north east);
    \end{tikzpicture}%
    \vspace*{0.5cm}%
    {\LARGE\bfseries\sffamily #1}%
  }

\titleformat{\subsection}
  {\normalfont\Large\bfseries\sffamily}
  {\thesubsection}
  {0.6em}
  {#1}

\titleformat{\subsubsection}
  {\normalfont\large\bfseries\sffamily}
  {\thesubsubsection}
  {0.6em}
  {\raisebox{-3pt}[0pt][0pt]{
    \smash{\hspace{-0.5em}
      \textcolor{colleaguepurple}
      {\rule{3pt}{\baselineskip}}\hspace{0.6em}}}#1}

\titlespacing*{\section}{0pt}{0pt}{24pt}
\titlespacing*{\subsection}{0pt}{20pt}{8pt}
\titlespacing*{\subsubsection}{0pt}{16pt}{6pt}
\KOMAoption{captions}{tableheading,figureheading}
\makeatletter
\@ifpackageloaded{caption}{}{\usepackage{caption}}
\AtBeginDocument{%
\ifdefined\contentsname
  \renewcommand*\contentsname{Table of contents}
\else
  \newcommand\contentsname{Table of contents}
\fi
\ifdefined\listfigurename
  \renewcommand*\listfigurename{List of Figures}
\else
  \newcommand\listfigurename{List of Figures}
\fi
\ifdefined\listtablename
  \renewcommand*\listtablename{List of Tables}
\else
  \newcommand\listtablename{List of Tables}
\fi
\ifdefined\figurename
  \renewcommand*\figurename{Figure}
\else
  \newcommand\figurename{Figure}
\fi
\ifdefined\tablename
  \renewcommand*\tablename{Table}
\else
  \newcommand\tablename{Table}
\fi
}
\@ifpackageloaded{float}{}{\usepackage{float}}
\floatstyle{ruled}
\@ifundefined{c@chapter}{\newfloat{codelisting}{h}{lop}}{\newfloat{codelisting}{h}{lop}[chapter]}
\floatname{codelisting}{Listing}

\makeatother
\makeatletter
\@ifpackageloaded{caption}{}{\usepackage{caption}}
\@ifpackageloaded{subcaption}{}{\usepackage{subcaption}}
\makeatother
\usepackage{bookmark}
\IfFileExists{xurl.sty}{\usepackage{xurl}}{} 
\hypersetup{
  pdftitle={Colleague AI Implementation Study:},
  colorlinks=true,
  linkcolor={blue},
  filecolor={Maroon},
  citecolor={Blue},
  urlcolor={Blue},
  pdfcreator={LaTeX via pandoc}}

\title{Colleague AI Implementation Study:}
\usepackage{etoolbox}
\makeatletter
\providecommand{\subtitle}[1]{
  \apptocmd{\@title}{\par {\large #1 \par}}{}{}
}
\makeatother
\subtitle{Cross-District Midyear Analysis}
\author{}
\date{}
\begin{document}
\includepdf[pages=1,fitpaper]{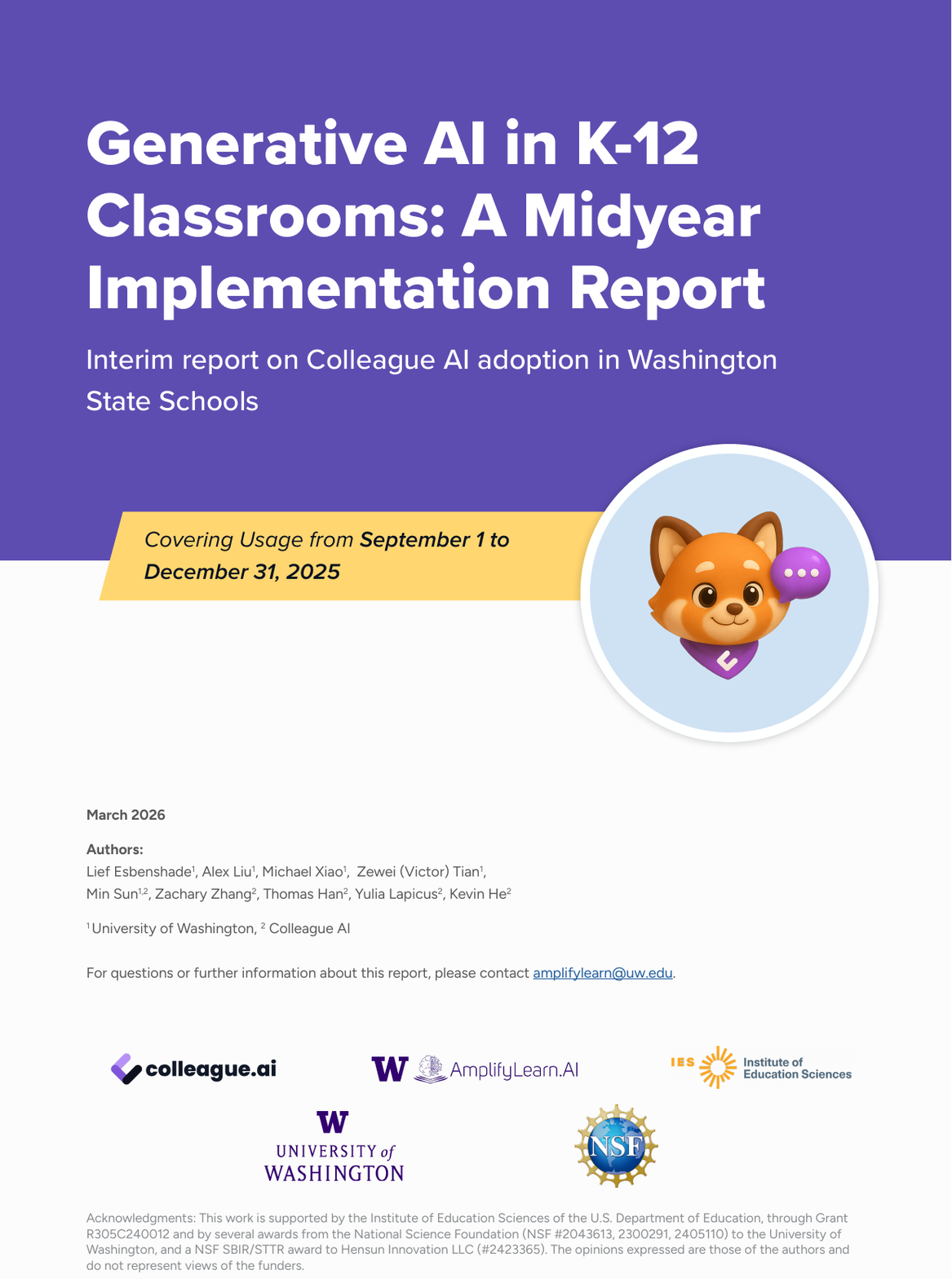}

\section{Executive Summary}\label{executive-summary}

This mid-year report summarizes teacher use of Colleague AI across 12
Washington State school districts from September 1 to December 31, 2025.
Produced jointly by Colleague AI and AmplifyLearn.AI at the University
of Washington, this report aggregates platform data and
district-provided administrative records to provide an early look at how
teachers engaged with AI during the first half of the 2025-26 school
year. The districts vary in size from small districts with a few
thousand students to large districts with up to thirty thousand
students. The districts are rural, suburban, and urban. Only a subset of
districts were able to provide mid-year administrative data, and
findings that link teachers' use of Colleague AI to student
characteristics should be interpreted as preliminary signals.

\begin{keyfindings}

\subsection{Key Findings}\label{key-findings}

\begin{itemize}
\item
  \textbf{Adoption grew steadily over the fall.} Across 12 participating
  districts, 1,438 teachers used Colleague AI in September and a total
  of 2,891 teachers had used Colleague AI by the end of December,
  exchanging over 412,304 messages with the AI assistant. We have
  observed continued growth in platform adoption; as of March 31, 2026
  adoption across these districts has increased by 61\% to 4,644
  teachers.
\item
  \textbf{Among active users, teachers averaged 12 sessions and 3.7
  hours on the platform.} Per session, teachers spent about 18 minutes.
  Teachers with more sessions also tended to have longer sessions and
  send more messages.
\item
  \textbf{Claire AI and lesson planning dominated usage.} Conversations
  with the Claire AI assistant accounted for the large majority of
  platform activity, and lesson planning features made up 60-99\% of
  usage in every participating district. Frequent users concentrated
  even more heavily on Claire, while lighter users and later adopters
  explored a broader mix of features.
\item
  \textbf{Teachers serving different student populations engaged with AI
  differently.} While overall usage intensity was similar across
  classroom contexts, teachers serving higher shares of English Language
  Learners or Special Education students referenced Student Profiles and
  Differentiation \& Accessibility more often in their requests.
  Teachers serving more SPED students and those with wider variability
  in their students' test scores also made relatively more use of Lesson
  Delivery features like Generate Image and Generate Interactive.
\item
  \textbf{Small, positive relationship between Teacher AI Use and
  Student ELA scores.} Examining 12,005 students with matched beginning-
  and middle-of-year test scores, we found that average tested students'
  teachers used Colleague AI for approximately 3 hours, and this use was
  associated with a statistically significant 0.025 standard deviation
  increase in reading interim test scores; the corresponding association
  in math was not statistically distinguishable from zero. This is a
  preliminary correlation, and the small effect size is an encouraging
  signal for AI adoption.
\end{itemize}

\end{keyfindings}

\newpage{}

Analysis of pedagogical topics in teacher request messages to the AI
platform found that across all districts, Planning and Explicit Teaching
were the most common pedagogical topics in teacher requests, followed by
Assessment, Critical Thinking and Inquiry, and Student Profiles.
Non-educational conversations were rare. Comparing teacher requests to
AI responses, we found that AI responses consistently surfaced
additional pedagogical topics---particularly around Feedback, Student
Profiles, and Differentiation---suggesting that Colleague AI is not only
responding to what teachers ask but also broadening the pedagogical
scope of the conversation. The one topic more prevalent in teacher
messages than in AI responses was Discourse Continuity, reflecting
teachers' active role in following up, refining, and redirecting AI
output across conversational turns.

Teachers serving different student populations used Colleague AI in
distinct ways. While overall usage intensity was similar across teachers
regardless of the proportion of English Language Learners (ELL) or
Special Education (SPED) students in their classrooms, the features and
topics teachers engaged with differed meaningfully. Teachers serving
higher shares of ELL students referenced Student Profiles,
Differentiation and Accessibility, Engagement and Motivation, and
Project-Based Learning more frequently in their requests. Teachers
serving more SPED students similarly emphasized Student Profiles and
Differentiation, and were notably more likely to use Lesson Delivery
features such as Generate Image and Generate Interactive. Teachers whose
classrooms had lower average test scores, or wider score spreads, were
also more likely to reference Differentiation, Student Profiles, and
Engagement, pointing to AI as a potential resource for supporting
instructional differentiation in heterogeneous classrooms.

Preliminary analysis links teacher AI use to modest gains in reading
scores. Analyzing 12,005 students, we examined whether students whose
teachers used Colleague AI more intensively had higher mid-year scores,
controlling for beginning of year scores, student characteristics, and
assessment and district fixed effects. In reading, three hours of
teacher AI use was associated with a statistically significant increase
of approximately 0.025 standard deviations in mid-year scores, a result
that held across model specifications and alternative measures of AI
engagement. The corresponding association in math was not statistically
distinguishable from zero. These findings are correlational and reflect
only the first four months of a voluntary adoption period, but this
offers an encouraging early signal. We note that teacher adoption of
Colleague AI was voluntary, and we may be observing a selection effect
where the teachers using Colleague AI differ from other teachers in
important ways that also influence student learning.

Districts included in this report agreed to participate in the study
prior to the beginning of the 2024-25 school year and completed Data Use
Agreements with Colleague AI confirming their participation. Among
these, some districts paused their AI implementation plans during the
study period or elected to pilot alternative AI tools, resulting in
variable levels of platform engagement. The districts included in this
study represent a small subset of Colleague AI's overall user base.
Individual districts are not identified in this report.

This research has been approved by the University of Washington
Institutional Review Board. Colleague AI has provided training resources
and professional development opportunities for teachers, all teacher use
of Colleague AI is voluntary. All data analyzed here has been
de-identified.

Several of the twelve participating districts provided mid-year
administrative records. These records enable analyses of usage patterns
by student demographics and student achievement. These supplementary
analyses are presented in later sections with sample sizes noted (see
\hyperref[sec-appendix-datasources]{Appendix: Data Sources} for the
breakdown of data availability by file type.)

\textbf{For questions about this report please contact
amplifylearn@uw.edu.}

\newpage{}

\section{Teacher Usage Levels}\label{teacher-usage-levels}

\subsection{Usage of Colleague AI Increased Steadily Over the
Fall}\label{usage-of-colleague-ai-increased-steadily-over-the-fall}

2891 teachers used Colleague AI as of December 31, 2025, representing
nearly 30\% of the teachers in the participating districts. For context,
a nationally representative survey of K--12 science and mathematics
teachers found that in May 2025 approximately half of teachers reported
using generative AI tools, and most of those (44\% of surveyed teachers)
reported using ChatGPT rather than an education-specific platform
(Esbenshade et al., 2025). In four of the districts participating in
this study, teacher use of Colleague AI surpassed the national average
rate of ChatGPT usage.\footnote{\emph{A note on the participation
  denominator:} Platform rostering data reflects all accounts
  provisioned by the district on Colleague AI, which in many cases
  substantially exceeds the number of active classroom teachers. In
  districts that did not configure automated rostering, no denominator
  is possible as all observed accounts were manually created by
  individual teachers. We use the National Center for Education
  Statistics (NCES) full-time equivalent (FTE) teacher count for each
  district to calculate the participation rate consistently across
  districts. While NCES figures may --- as an FTE value --- undercount
  individuals, we believe they offer a more consistent estimate of the
  teacher adoption rate than the roster counts.}.

\subsubsection{Platform Adoption: Cumulative teacher engagement and
overall participation
rate}\label{platform-adoption-cumulative-teacher-engagement-and-overall-participation-rate}

\pandocbounded{\includegraphics[keepaspectratio]{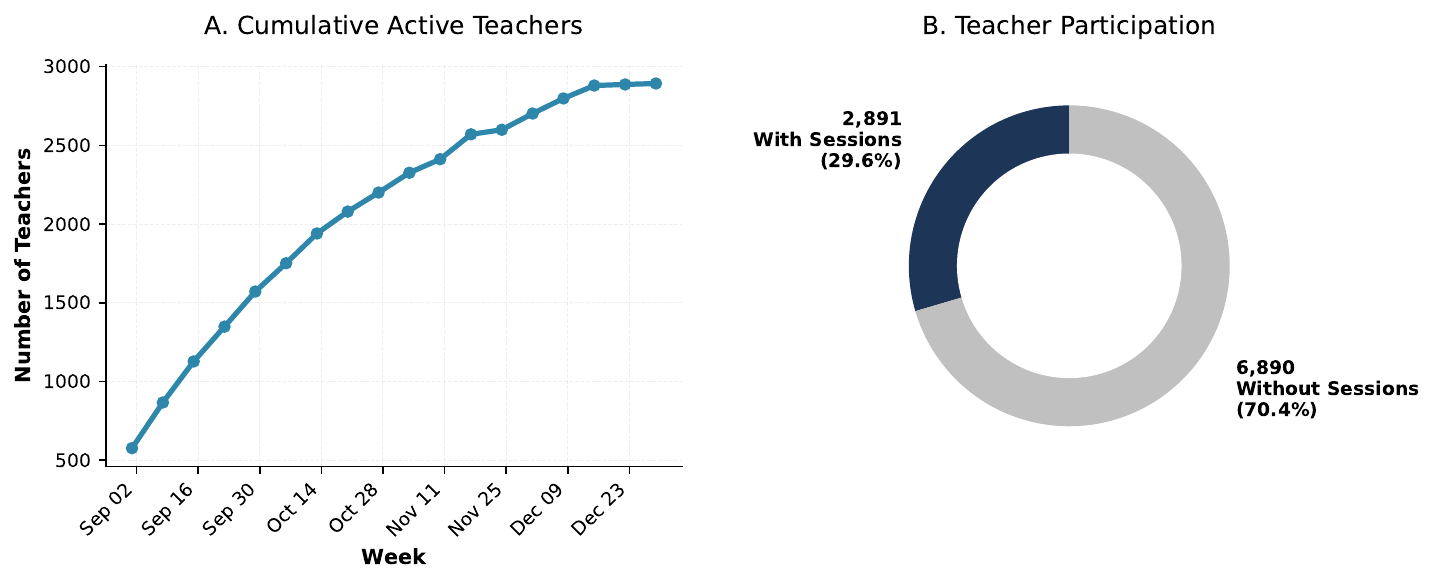}}

\begin{figure}

\begin{minipage}{0.50\linewidth}

\newpage{}

\end{minipage}%
\begin{minipage}{0.50\linewidth}

\subsubsection{Participation Rate by
District}\label{participation-rate-by-district}

\pandocbounded{\includegraphics[keepaspectratio]{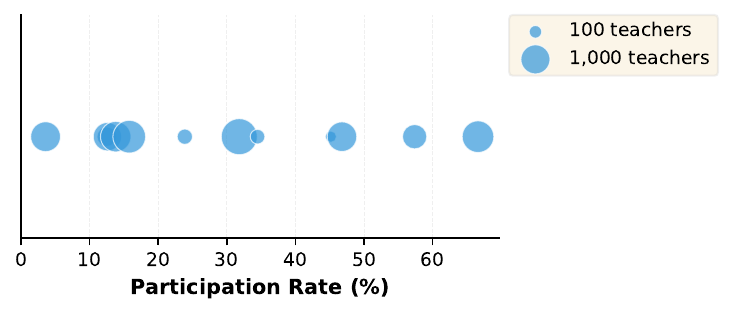}}

\end{minipage}%

\end{figure}%

We also compare the adoption rate across districts. In this dot plot,
each point represents one participating district's adoption rate as of
December 31, 2025. We observed a substantial spread in teacher
engagement during the fall across school districts. There is not a clear
relationship between district size and participation rate. This report
on Colleague AI's adoption does not just cover ideal use with invested
school partners, instead we include a wide range of districts in order
to best understand how an AI platform is broadly used by teachers.

\subsection{We measure teacher usage of Colleague AI in four
ways.}\label{we-measure-teacher-usage-of-colleague-ai-in-four-ways.}

\begin{enumerate}
\def\labelenumi{\arabic{enumi}.}
\tightlist
\item
  \textbf{Sessions} captures the number of times teachers use Colleague
  AI. We define a session as beginning with the first AI interaction and
  ending after 60 minutes of inactivity.
\item
  \textbf{Session Minutes} records the total time a teacher spent
  interacting with AI features, measured from the first to last observed
  interaction within each session. It does not include the 60 minutes of
  inactivity that signal the end of a session.
\item
  \textbf{AI Conversations} measures the number of chat threads a
  teacher created. The same thread can be returned to across sessions,
  and a teacher may create multiple conversations within a single
  session.
\item
  \textbf{AI Messages} measures the total number of messages teachers
  exchanged with the AI platform.
\end{enumerate}

In the four histograms that follow, each histogram shows the per-teacher
distribution for one usage measure; teachers with zero interactions are
excluded, and values are truncated at the 99th percentile. Teachers on
average had 12.0 sessions, spent 3.7 hours on the platform, had 12.0
conversations, and sent 143 messages. Across all four measures, most
teachers used Colleague AI lightly, while a smaller group became
frequent users. Sessions, minutes, conversations, and messages are
generally well correlated: teachers with more sessions tend to also
spend more time and send more messages, though some teachers favor a few
long sessions while others prefer many short ones. Total n sizes vary
slightly across the four figures due to technical variation in how
different platform activities are logged. The n-size for conversations,
in particular, is lower as that usage measure is only present for
teachers who interacted with the Claire AI conversational feature.

\newpage{}

\subsubsection{Teacher Usage by Sessions, Minutes, Conversations, and
Messages}\label{teacher-usage-by-sessions-minutes-conversations-and-messages}

\begin{figure}

\begin{minipage}{0.50\linewidth}

\pandocbounded{\includegraphics[keepaspectratio]{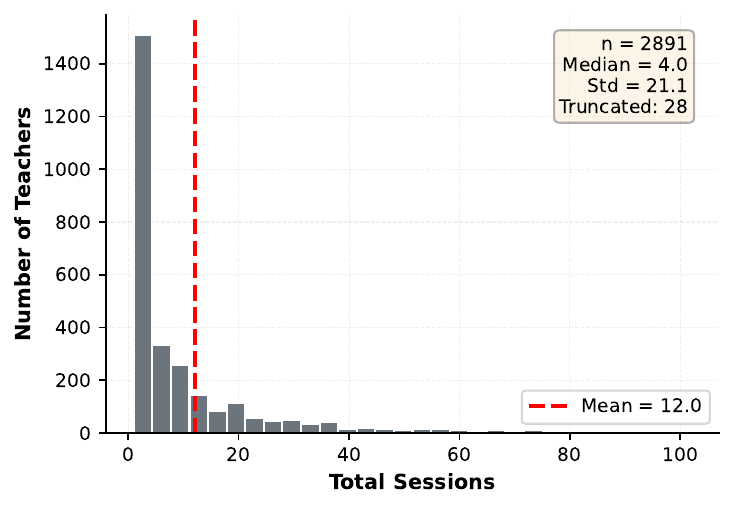}}

\end{minipage}%
\begin{minipage}{0.50\linewidth}

\pandocbounded{\includegraphics[keepaspectratio]{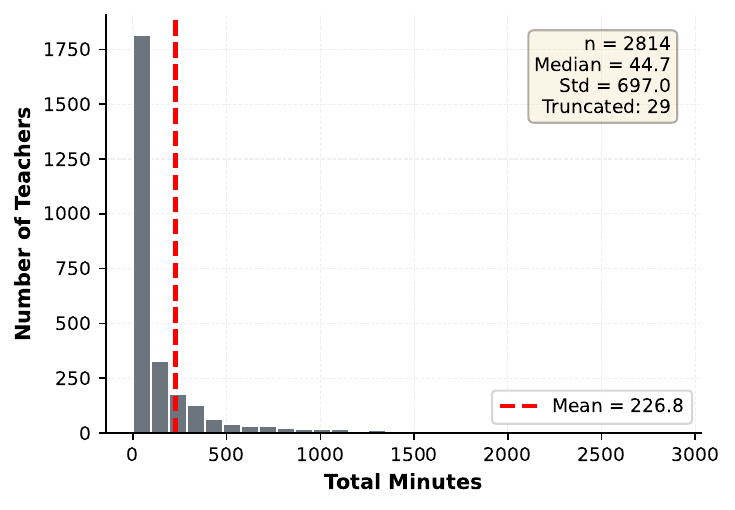}}

\end{minipage}%
\newline
\begin{minipage}{0.50\linewidth}

\pandocbounded{\includegraphics[keepaspectratio]{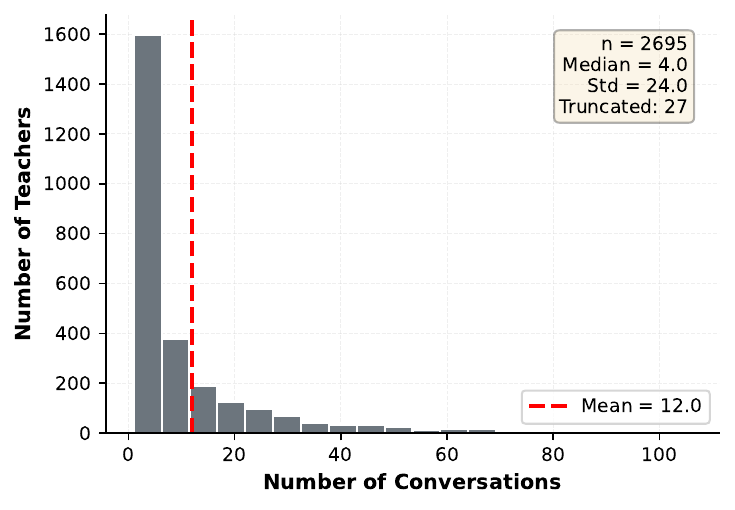}}

\end{minipage}%
\begin{minipage}{0.50\linewidth}

\pandocbounded{\includegraphics[keepaspectratio]{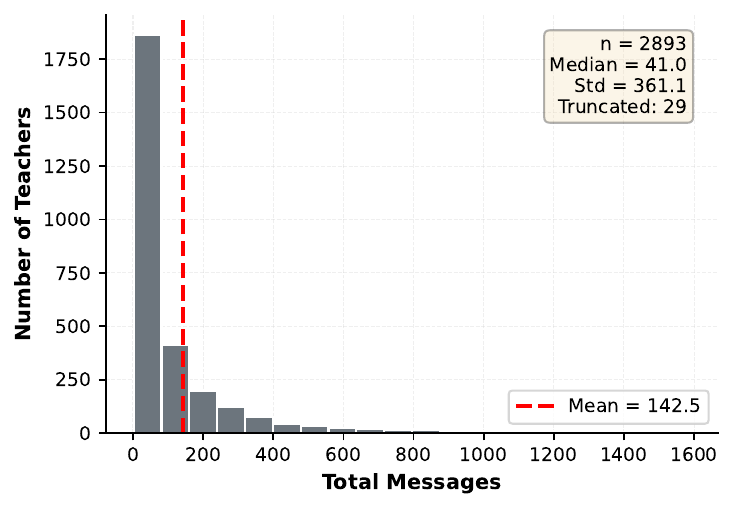}}

\end{minipage}%

\end{figure}%

\subsubsection{Correlation Between Usage
Measures}\label{correlation-between-usage-measures}

\pandocbounded{\includegraphics[keepaspectratio]{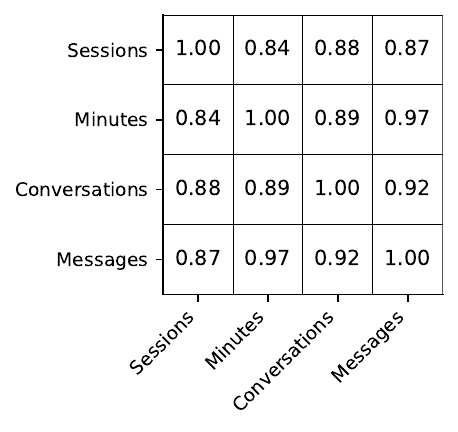}}

\newpage{}

The four measures are strongly correlated, and for clarity we focus on
total sessions as the primary usage metric in the analyses that follow.

\subsubsection{Sessions per Teacher by
District}\label{sessions-per-teacher-by-district}

\pandocbounded{\includegraphics[keepaspectratio]{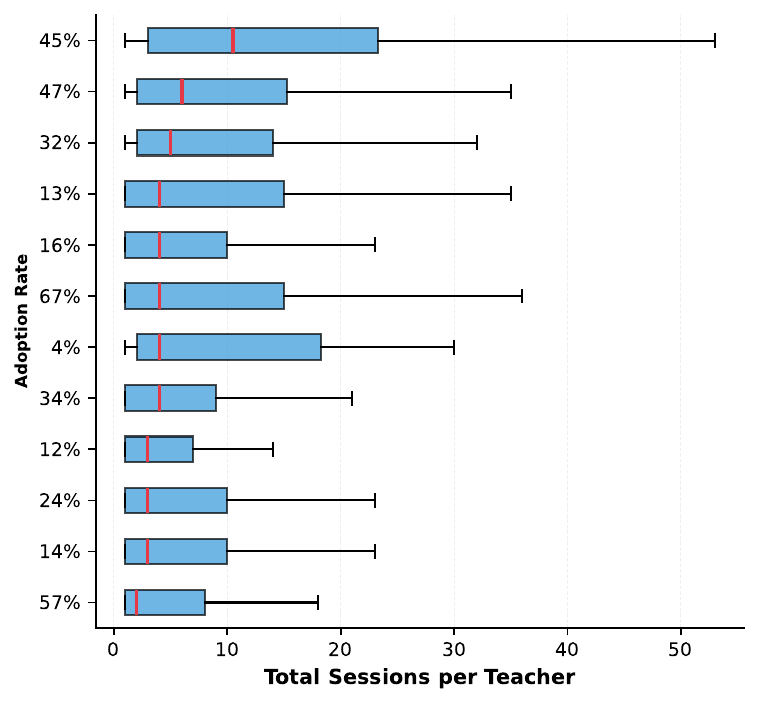}}

This boxplot shows the distribution of total sessions among active
teachers by district. The districts are labelled based on their overall
adoption rate. From this figure we can see that there is not a clear
relationship between the overall adoption rate and the intensity of
adoption as measured by per teacher sessions. Two districts with near
50\% adoption rates have the highest per teacher median session rates,
but the district with a 57\% adoption rate has the lowest median per
teacher session rate. The graph is sorted in descending order based on
the median teacher in each district, but we see substantial variation at
the third quartile, with some districts that have lower median use
showing substantially higher third quartile use. All of this is to say
that adoption of Colleague AI in districts is complex, with substantial
variation in how widespread and how intensive usage is.

\newpage{}

To look more closely at who uses the platform and how, we define two
groupings:

\begin{enumerate}
\def\labelenumi{\arabic{enumi}.}
\tightlist
\item
  \textbf{Usage level} based on total sessions, which captures how many
  distinct times a teacher engaged with the platform, we define a light
  user group that has tried Colleague AI 1 to 4 times, a moderate user
  group who has used Colleague AI 5 to 20 times, and a frequent user
  group that has used Colleague AI 21 or more times. The frequent user
  group averaged over one session a week during the study period.
\item
  \textbf{Adoption cohort} compares early adopters with those who waited
  before trying Colleague AI. We define a September cohort of teachers
  who used Colleague AI in September 2025 and contrast it to an October+
  cohort whose first recorded session occurred after October 1st, 2025.
\end{enumerate}

\begin{figure}

\begin{minipage}{\linewidth}

\subsubsection{Teacher Usage Groups and Adoption
Cohorts}\label{teacher-usage-groups-and-adoption-cohorts}

\pandocbounded{\includegraphics[keepaspectratio]{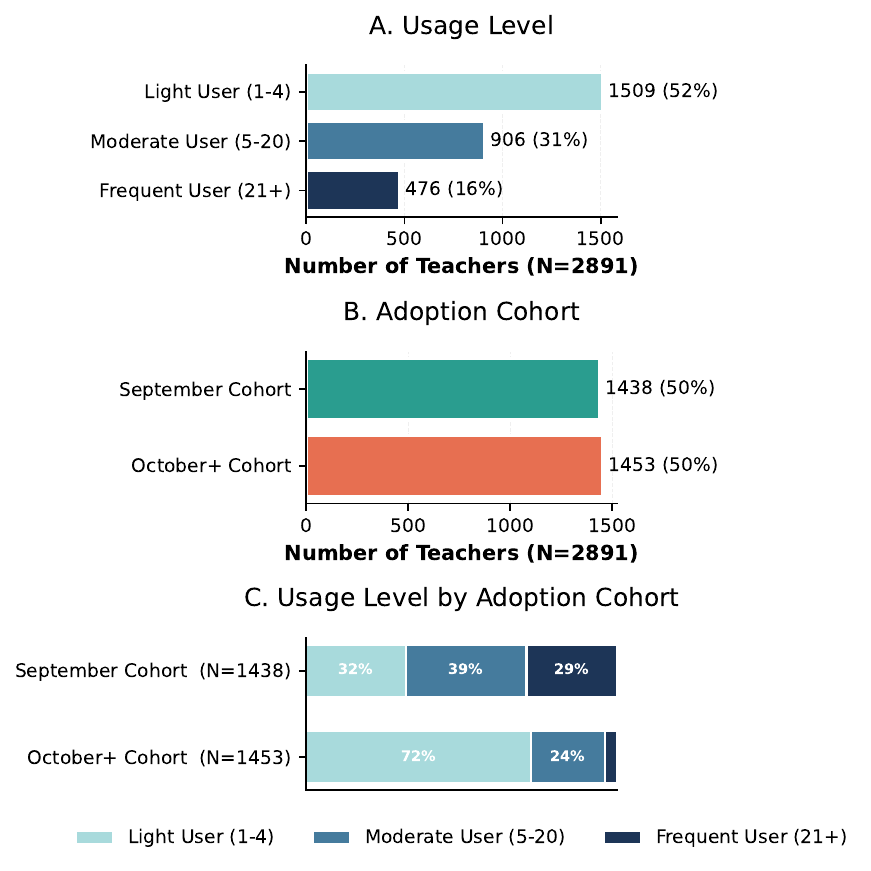}}

\end{minipage}%

\end{figure}%

\newpage{}

\section{How Teachers Use The
Platform}\label{how-teachers-use-the-platform}

Colleague AI offers a range of features, which we group into three
categories (see \hyperref[sec-appendix-features]{Appendix: Explanation
of Feature Categories} for full definitions):

\begin{itemize}
\tightlist
\item
  \textbf{Lesson Planning} encompasses features meant to aid in
  preparing lessons and includes: conversations with the AI assistant
  Claire, lesson plan creation, and the lesson plan quality measure
  tool.
\item
  \textbf{Lesson Delivery} includes features designed to create
  materials to support teaching like: image, interactive, slide,
  podcast, and diagram generation.
\item
  \textbf{Student Interaction and Feedback} groups together features
  that interface more directly with students such as: rubric generation,
  AI grading, and student-facing tools like the AI Tutor.
\end{itemize}

\subsection{Teachers Mostly Use Lesson Planning
Features}\label{teachers-mostly-use-lesson-planning-features}

These categories may overlap in practice; conversations with Claire AI
in Brainstorm Ideas, in particular, can span planning, delivery, and
feedback. Across all districts, Claire is by far the most used feature
on the platform.

The figures that follow shows each district's share of feature usage
across the three categories and the breakdown of features used within
the categories. While lesson planning dominates in every district,
districts vary in how much they use lesson delivery and student
interaction features.

Conversations with Claire AI are far and away the most common use of the
platform. As such, the lesson planning category predominates. We observe
that frequent users concentrated on lesson planning features, while
lighter users and later adopters tended to try a broader mix. This
pattern indicates that as users spend more time on the platform, more of
the usage is concentrated on conversations with Claire AI.

Colleague AI is continuing to develop the platform. In early 2026,
improvements were released to Generate Slides and a new AI Grading
workflow was added to the Claire chat interface. Further refinements
including a live-chat mode are also being tested for the student facing
features. In addition, with advances in AI tool calling and agentic
capabilities, Colleague AI is integrating more features directly into
the Claire AI chat interface. The end-of-year report will look at
adoption of these updates.

\begin{figure}

\begin{minipage}{\linewidth}

\subsubsection{Feature Usage Overview}\label{feature-usage-overview}

\pandocbounded{\includegraphics[keepaspectratio]{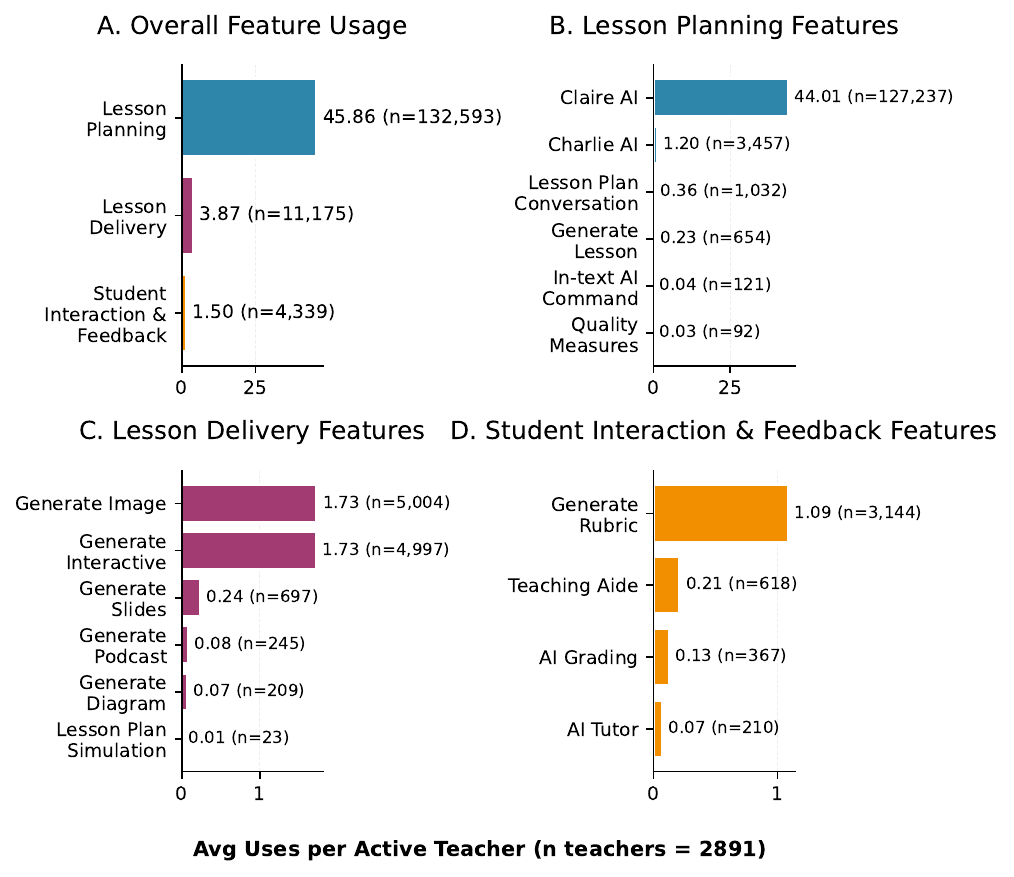}}

\end{minipage}%

\end{figure}%

\subsubsection{Feature Category Mix by
District}\label{feature-category-mix-by-district}

\pandocbounded{\includegraphics[keepaspectratio]{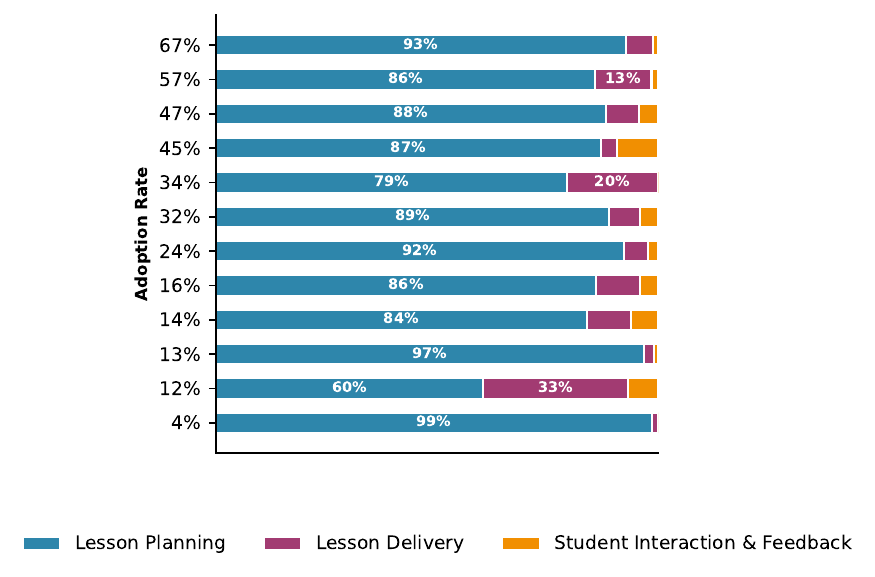}}

\subsubsection{Features by Teacher
Groups}\label{features-by-teacher-groups}

\pandocbounded{\includegraphics[keepaspectratio]{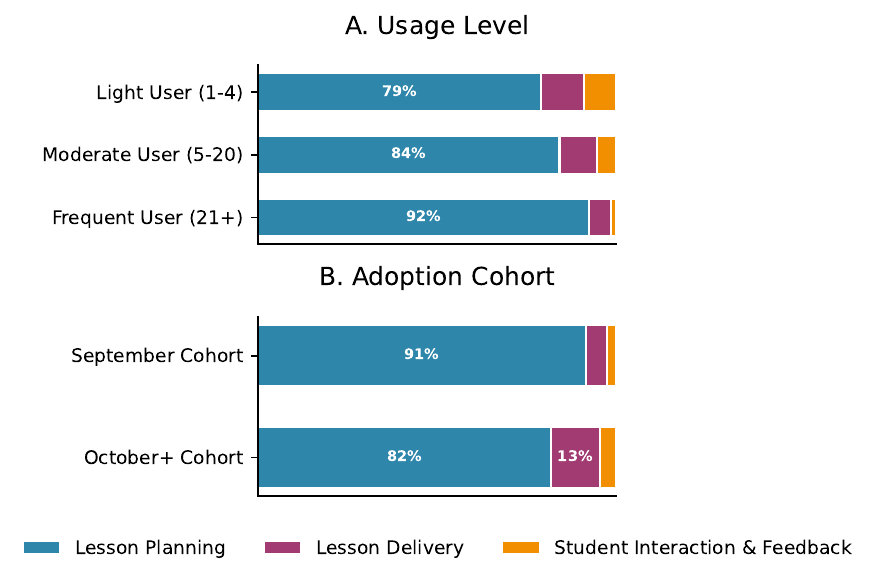}}

\newpage{}

\subsection{Pedagogical Topics of Teacher-AI
Conversations}\label{pedagogical-topics-of-teacher-ai-conversations}

To further understand how teachers use Colleague AI, the platform
automatically tags the pedagogical topics present in each conversation.
At the end of this report we have included an
\hyperref[sec-appendix-topics]{appendix} with detailed definitions of
each pedagogical topic\footnote{More information on how conversation
  topics are identified can be found in Liu, A., Esbenshade, L., Sarkar,
  S., Tian, V., Zhang, Z., He, K., \& Sun, M. (2025). How K-12 Educators
  Use AI: LLM-Assisted Qualitative Analysis at Scale. arXiv preprint
  arXiv:2507.17985}.

We begin by comparing across all the Teacher-AI conversations the
relative prevalence of pedagogical topics in the message requests sent
by teachers to those in the AI responses. The AI responses consistently
add pedagogical topics to the conversation. Across districts, planning
and explicit teaching, topics which refer respectively to prototyping
lesson plans and explaining core content area topics, are the most
commonly referenced by teachers. Non-educational conversation topics are
very rare. Discourse Continuity refers to cross-turn coherence in
conversational threads and the extent to which messages refer back to,
and build upon, prior messages. This is the primary topic that is more
prevalent in teacher messages than in AI assistant messages.

The following charts examine the pedagogical topics in teachers'
requests to Colleague AI, showing how different groups of teachers
interact with the platform.

\subsubsection{AI Conversations Surface More Pedagogical
Discourse}\label{ai-conversations-surface-more-pedagogical-discourse}

\pandocbounded{\includegraphics[keepaspectratio]{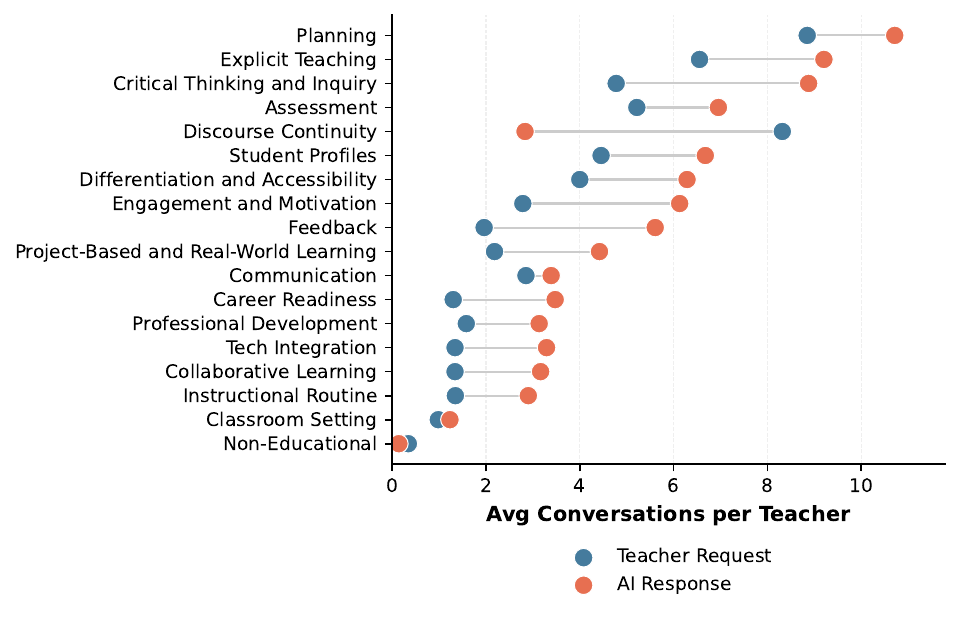}}

\subsubsection{Teacher Requests' Topic Distribution Across
Districts}\label{teacher-requests-topic-distribution-across-districts}

\pandocbounded{\includegraphics[keepaspectratio]{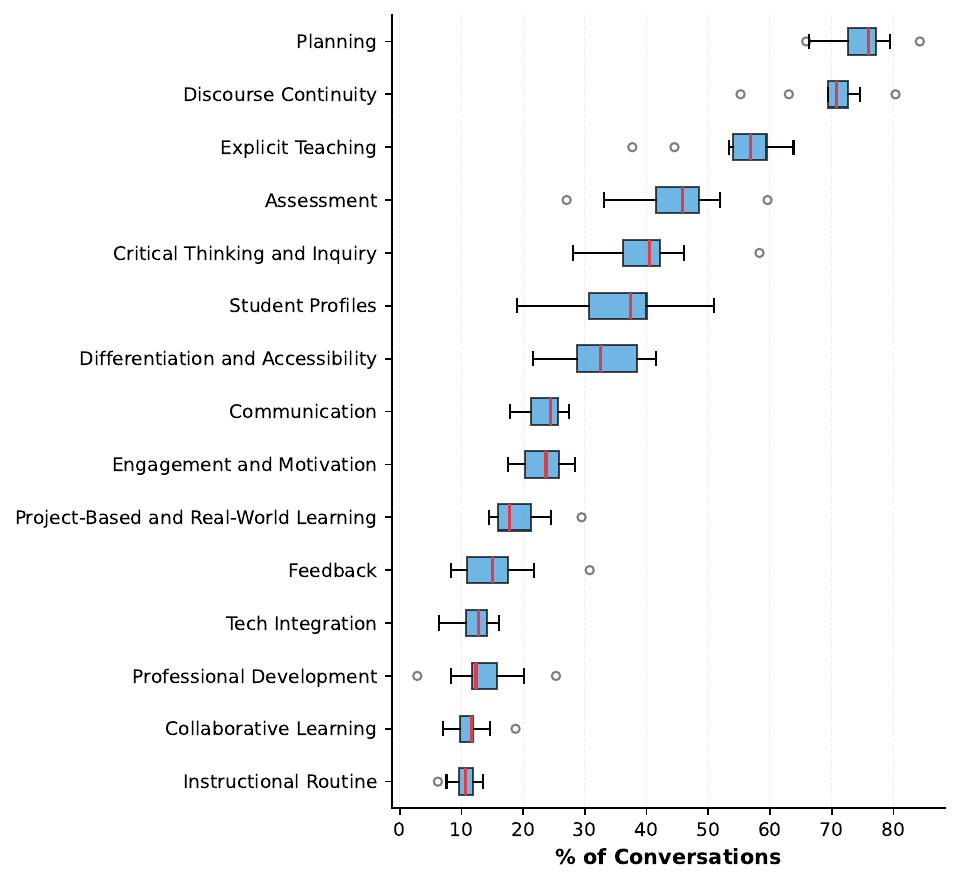}}

Each box summarizes one topic's prevalence across participating
districts. The Student Profiles topic, which refers to the explicit
description of specific types of student groups that are being taught,
has the widest range in use across districts. Differentiation \&
Accessibility, Critical Thinking \& Inquiry, and Assessment also saw
relatively wide ranges in prevalence across districts.

Relative to light or moderate users, frequent users of Colleague AI were
more likely to follow up on AI responses in subsequent messages
(Discourse Continuity) and more likely to use the AI tools to draft
messages for communicating with families or colleagues (Communication).

\subsubsection{Relative Prevalence of Pedagogical Topics (Teacher
Requests
Only)}\label{relative-prevalence-of-pedagogical-topics-teacher-requests-only}

\pandocbounded{\includegraphics[keepaspectratio]{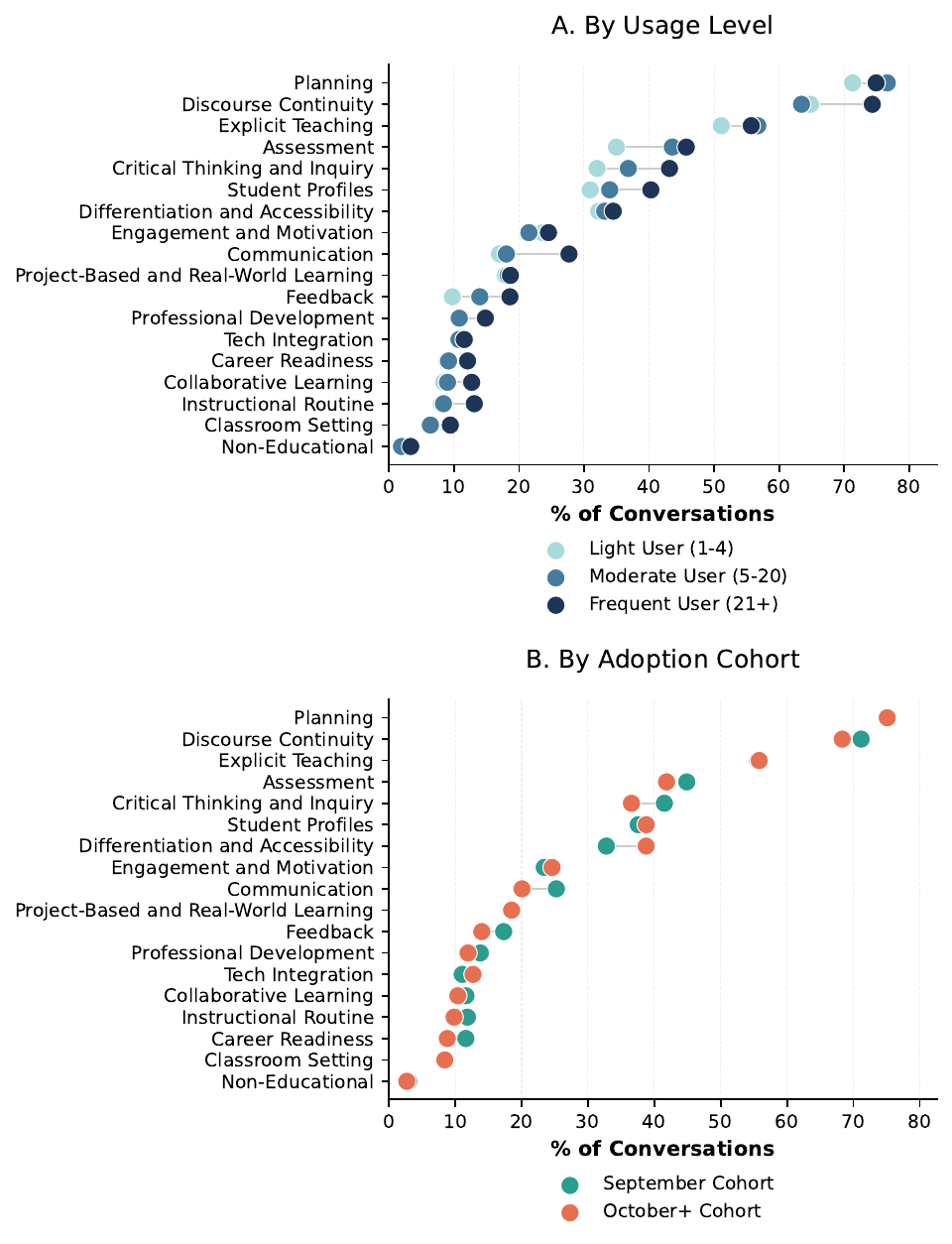}}

\newpage{}

\section{Student-Teacher Rostering
Linkage}\label{student-teacher-rostering-linkage}

Teachers are linked to students through classroom rosters on the
Colleague AI platform. Classroom records with associated student and
teacher accounts are created automatically through syncing with
platforms like Classlink and Clever. The rostered classes automatically
include some subject area information. Currently we observe that the
most common class subject label is ``other'' and will continue to work
to improve classroom subject mapping for analysis.

These linkages underpin the student-level analyses in this report. The
table below pools roster data across all participating districts. We
exclude classrooms with more than 70 students (the 95th percentile of
all classroom sizes) as likely representing digital classrooms that do
not directly relate to traditional classrooms.

We include only roster records that have a system update date on or
after July 01, 2025.

\begin{figure}

\begin{minipage}{0.50\linewidth}

\begin{longtable}[]{@{}lr@{}}
\toprule\noalign{}
Total Records & \\
\midrule\noalign{}
\endhead
\bottomrule\noalign{}
\endlastfoot
Teachers & 18,112 \\
Active teachers & 2,891 \\
Students & 77,486 \\
Classrooms & 28,400 \\
\end{longtable}

\end{minipage}%
\begin{minipage}{0.50\linewidth}

\begin{longtable}[]{@{}lr@{}}
\toprule\noalign{}
Averages & \\
\midrule\noalign{}
\endhead
\bottomrule\noalign{}
\endlastfoot
Students per classroom & 26 \\
Teachers per classroom & 1.6 \\
Classrooms per teacher & 9.9 \\
Classrooms per student & 9.6 \\
\end{longtable}

\end{minipage}%

\end{figure}%

\textbf{Classrooms by Subject Area}

\begin{longtable}[]{@{}lr@{}}
\toprule\noalign{}
Subject Area & Classrooms \\
\midrule\noalign{}
\endhead
\bottomrule\noalign{}
\endlastfoot
Other & 12,753 \\
ELA & 3,185 \\
Math & 3,032 \\
Science & 2,378 \\
Social Studies & 2,130 \\
Arts & 1,760 \\
PE & 1,381 \\
Computer Science \& Technology & 960 \\
Career \& Technical Education & 572 \\
English Language Learners & 235 \\
Special Education & 14 \\
\end{longtable}

\newpage{}

\section{Teacher Usage by Students
Taught}\label{teacher-usage-by-students-taught}

With district provided student demographic data, we examine how the
student populations teachers serve relate to their platform use.
Generative LLMs have potential to assist teachers in differentiating
their instructional materials to serve a broad range of students. In
this section we examine early signals in how teachers serving different
student populations are interacting with Colleague AI. Teachers are
linked to students through classroom rosters on the Colleague AI
platform; from these linkages we compute the percentage of each
teacher's students who are English Language Learners (ELL) or receive
Special Education (SPED) services.

Several patterns emerge. The overall usage level distribution is similar
across teachers with different student populations. Approximately half
are light users, 30\% are moderate users, and 20\% are frequent users.
We see larger differences in terms of features used. Teachers serving
more English Language Learners than the state average were less likely
to use Lesson Delivery features. On the other hand, teachers serving
more Special Education students than the state average were more likely
to use Lesson Delivery features. The most commonly used Lesson Delivery
features include Generate Image and Generate Interactive. In future work
we will investigate these features more deeply to understand differences
in adoption between teachers serving Special Education students and
English Language Learners.

Teachers also cover different pedagogical topics depending on their
student demographics. Teachers with more English Language Learners are
more likely to reference Student Profiles and Differentiation \&
Accessibility topics, as well as Engagement \& Motivation and
Project-Based \& Real-World Learning. Similarly, teachers with more
Special Education students are more likely to reference Student Profiles
and Differentiation \& Accessibility in their AI requests, but do not
have increased prevalence of conversations relating to Engagement or
Project-Based Learning. We will continue to monitor these trends and in
future work will expand on this analysis.

\subsection{Data linkage and
classification}\label{data-linkage-and-classification}

We are able to link over 500 teachers who used Colleague AI to their
classroom students. Teachers are grouped by the proportion of English
Learner and Special Education students in their classrooms, using
categories anchored to Washington state enrollment patterns. These
groupings describe the \emph{teaching context}, the instructional
environment in which a teacher works, rather than characterizing
students or implying judgments about classroom quality.

For \textbf{English Language Learner} classification, we use the term
and definition consistent with the Washington School Improvement
Framework (WSIF) accountability reporting. Categories are anchored to
Washington's statewide rate of approximately 12\% of public school
students receiving Transitional Bilingual Instruction Program (TBIP)
services (OSPI, 2025). Teachers with fewer than 8\% English Learner
students are classified as below the state average; those between 8\%
and 16\% are near the state average; and those above 16\% serve English
Learner populations substantially above the state norm, a threshold at
which research suggests instructional demands shift meaningfully
(Turgut, Sahin, \& Huerta, 2016).

For \textbf{Special Education} classification, categories are anchored
to state and national benchmarks. Nationally, 15\% of public school
students received services under IDEA in 2022--23, with Washington state
rates close to this figure (NCES, 2024). Teachers with fewer than 10\%
Special Education students are classified as below the state average;
those between 10\% and 20\% are near the state average; and those above
20\% serve Special Education populations well above the typical range
across states (Pew Research Center, 2023).

Student demographic data was provided by 4 of the 12 participating
districts, covering 45,956 students linked to 2,224 teachers. Of these
teachers, 502 used Colleague AI at least once during the study period.

\subsubsection{Teacher Distribution by Student ELL and SPED
Populations}\label{teacher-distribution-by-student-ell-and-sped-populations}

\pandocbounded{\includegraphics[keepaspectratio]{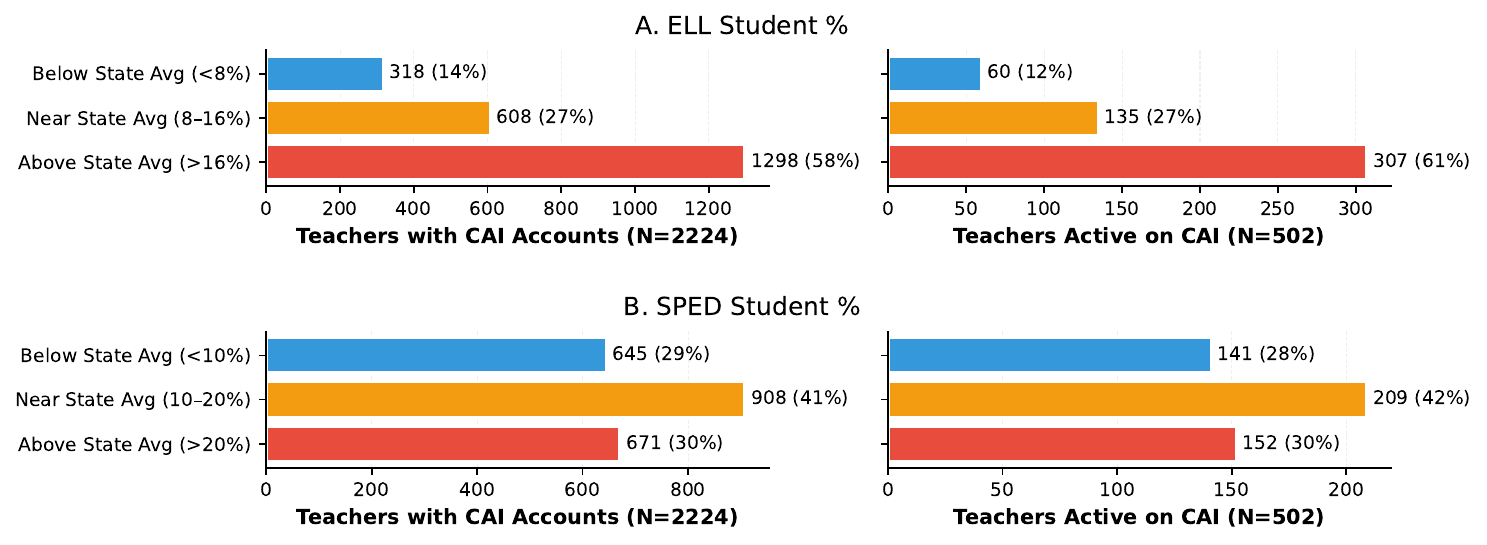}}

Among the teachers with linked student demographic data, we can compare
usage levels across teachers serving different proportions of ELL and
SPED students. We see that the overall distribution of teachers with
Colleague AI accounts with linked student data (left charts) is similar
to the distribution of teachers who used Colleague AI (right charts).
The adoption rate among teachers with linked student data is 22\%,
somewhat lower than in the overall study.

\subsubsection{Colleague AI Usage by ELL and SPED Student
Populations}\label{colleague-ai-usage-by-ell-and-sped-student-populations}

\pandocbounded{\includegraphics[keepaspectratio]{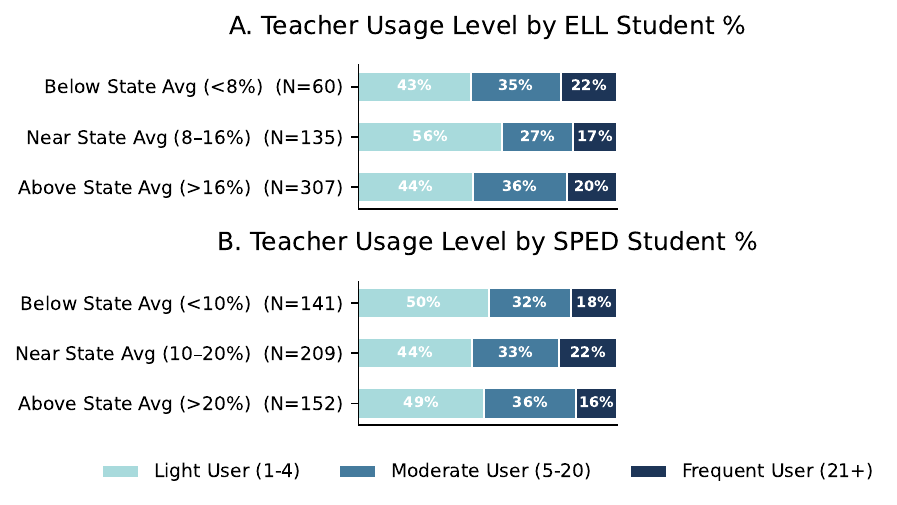}}

\subsubsection{Teacher Feature Use by ELL and SPED Student
Populations}\label{teacher-feature-use-by-ell-and-sped-student-populations}

\pandocbounded{\includegraphics[keepaspectratio]{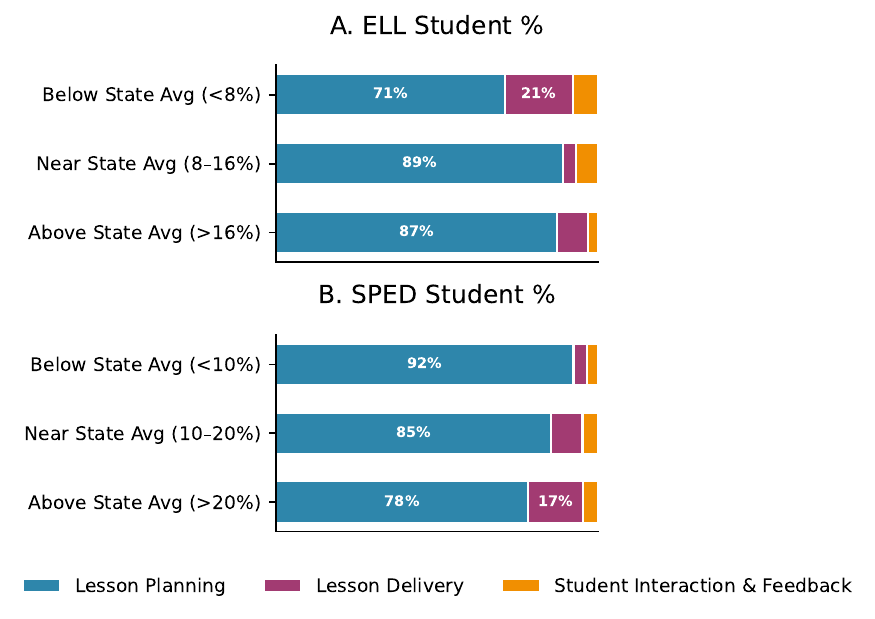}}

\subsubsection{Relative Prevalence of Pedagogical Topics by ELL and SPED
Student
Populations}\label{relative-prevalence-of-pedagogical-topics-by-ell-and-sped-student-populations}

\pandocbounded{\includegraphics[keepaspectratio]{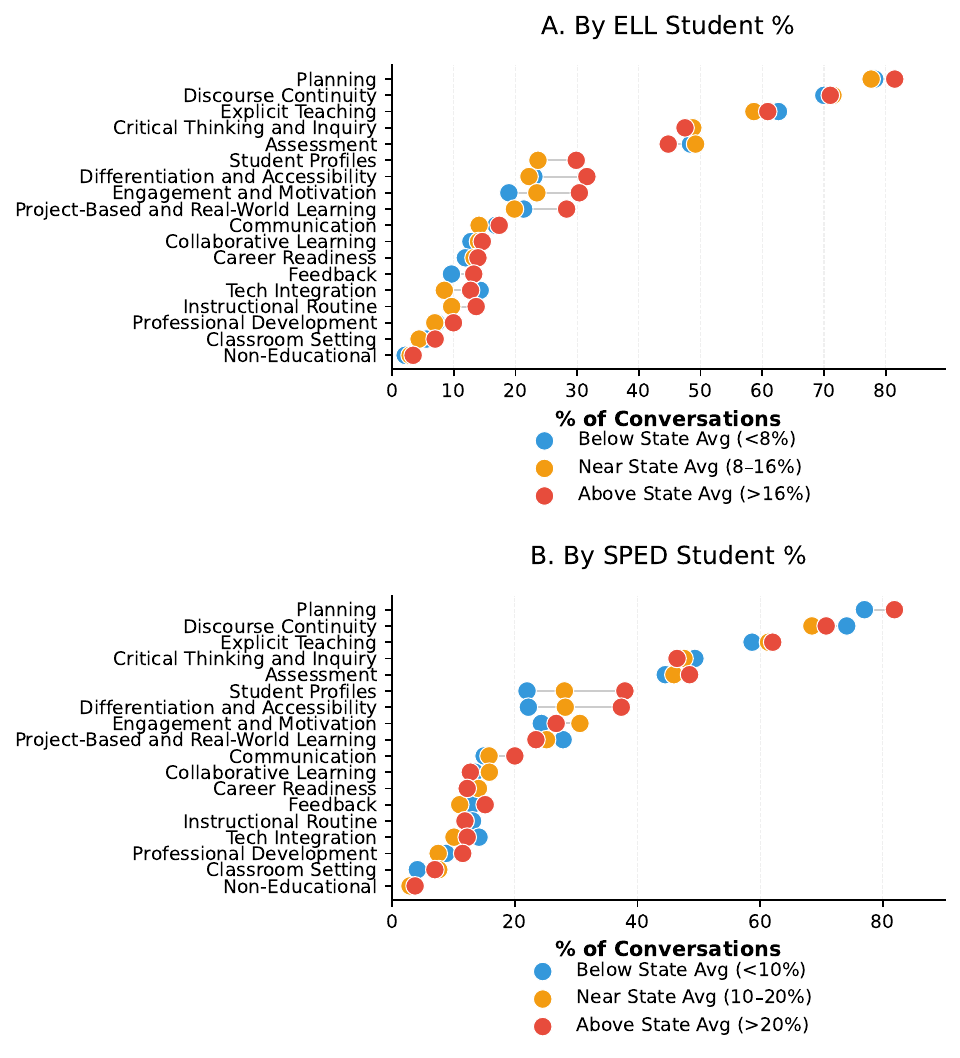}}

\newpage{}

\section{Teacher Usage by Student Assessment
Data}\label{teacher-usage-by-student-assessment-data}

With district provided student assessment data, we examine how teachers'
platform use relates to the academic achievement profiles of their
students. The results in this section describe the \emph{teaching
context} --- the classroom achievement profile a teacher works within
--- and are not a measure of teacher quality or of AI's contribution to
student learning. Continuing from the analysis in the previous section,
here we focus on the average score level among teachers' students, and
the degree of variability in the students' scores. After calculating the
average score among teachers' linked students, we group teachers into
thirds based on the overall distribution. We additionally disaggregate
teachers' usage of Colleague AI based on how wide the spread is between
the highest and lowest test scores of their students.

Within that frame, we see that teachers with lower-scoring classrooms
are slightly more likely to use Colleague AI; 37\% of teachers who use
Colleague AI have average classroom scores in the bottom 33\% of the
overall distribution. We also observe that 54\% of teachers with wide
score variability are light users of Colleague AI and only 13\% are
frequent users (vs 42\% light users and 23\% frequent users for teachers
with narrow score spreads). This has implications for further research
regarding AI adoption and use to support instructional differentiation.
We do see that teachers with wide test score variability did use Lesson
Delivery features relatively more often. Teachers with students in the
lower and middle third of the test score distribution were more likely
to reference Differentiation \& Accessibility, Student Profiles, and
Engagement \& Motivation in their AI requests. Patterns in pedagogical
topics are less clear based on test score variability. There is a
similar pattern with Differentiation, Student Profiles, and Engagement
as teachers with moderate or wide test score variability are more likely
to send message requests relating to those topics. There are also
indications that teachers with moderate, as opposed to narrow or wide
test score variability, were less likely to send message requests
referring to Assessment, which should be investigated further.

In early 2026 Colleague AI released significant upgrades to the AI
grading capabilities; these updates fall outside the scope of this
September -- December analysis. In future reports we will investigate
relationships between student achievement levels in classrooms and
teachers' adoption of AI grading and feedback functionality.

\subsection{Data linkage and
classification}\label{data-linkage-and-classification-1}

Student test scores are standardized (z-scored) within each combination
of test, subject, and grade level so that scores are comparable across
different assessments. Where multiple tests were available for a
district, each teacher's group assignment reflects a composite across
all available assessments rather than any single test. Teachers with
fewer than 5 linked students are excluded. Each teacher is characterized
along two dimensions based on their students' score distributions:

\begin{itemize}
\tightlist
\item
  \textbf{Mean Score Level} --- We calculate the average standardized
  test score of a teacher's linked students. Then, within each district
  we divide teachers into thirds grouping them in the Lower Third,
  Middle Third, or Upper Third. This is thus a relative, within-district
  measure to capture variability across teachers' classroom achievement
  levels. This is not a measure of teacher effectiveness but is a
  measure of the classroom context.
\item
  \textbf{Score Spread} --- Using the same teacher-student linkage, we
  calculate the difference between the lowest 10th percentile student's
  standardized test score and the highest 90th percentile student's
  standardized score. We then split this p10--p90 difference measure
  within each district into thirds, a Narrow, Moderate, and Wide
  grouping to measure the variability in achievement levels in teachers'
  classrooms. (See Reardon, 2011 for additional details about the
  p10--p90 measure.)
\end{itemize}

Student assessment data was provided by 3 of the 12 participating
districts, covering 172,655 test score records for 27,551 students.
Assessment data included a combination of WA State SBA assessments in
math and reading, WCAS science, STAR Reading and Math assessments, and
WIDA assessments. Scores were standardized within district, assessment,
grade, and subject. After linking through classroom rosters, 1,711
teachers had at least 5 students with scores, of whom 458 used Colleague
AI.

\subsubsection{Teacher Distribution by Student Score
Groups}\label{teacher-distribution-by-student-score-groups}

\pandocbounded{\includegraphics[keepaspectratio]{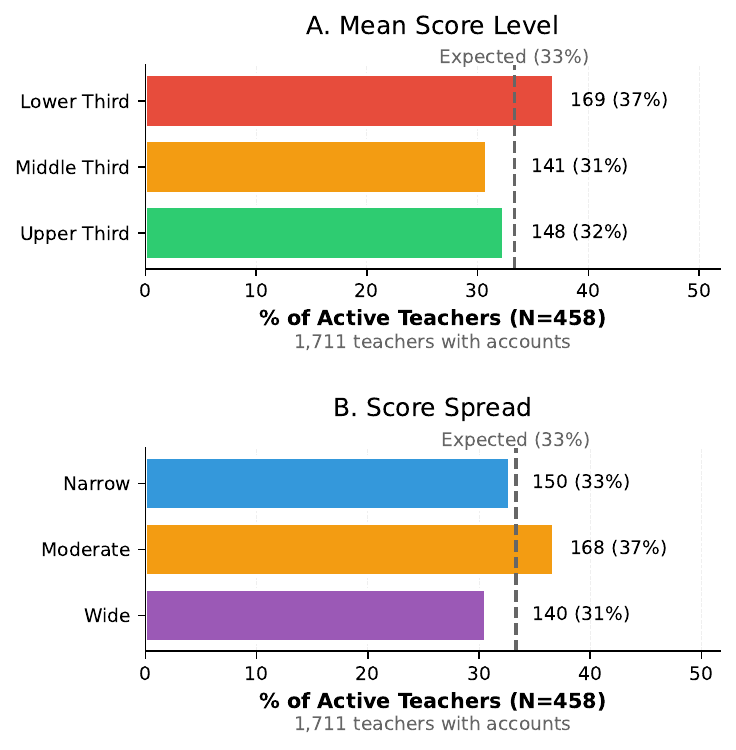}}

\subsubsection{Colleague AI Usage by Student Score
Groups}\label{colleague-ai-usage-by-student-score-groups}

\pandocbounded{\includegraphics[keepaspectratio]{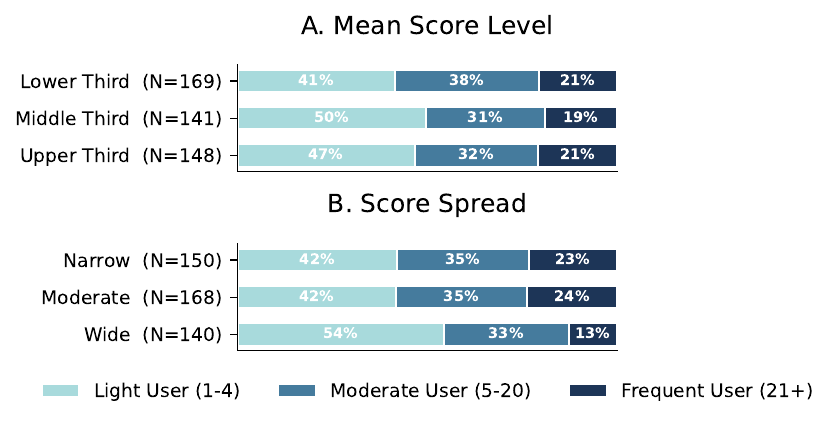}}

\subsubsection{Features by Student Score
Groups}\label{features-by-student-score-groups}

\pandocbounded{\includegraphics[keepaspectratio]{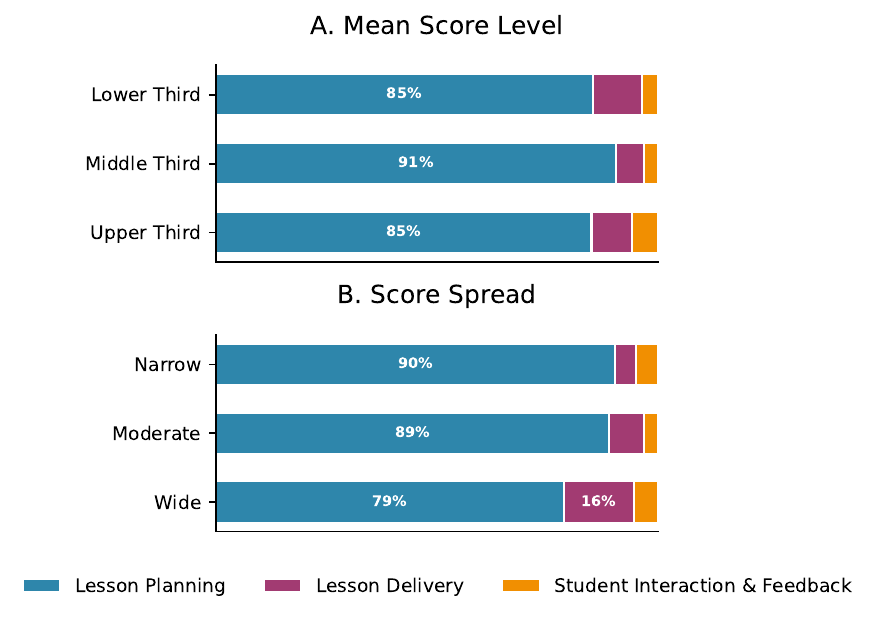}}

\subsubsection{Relative Prevalence of Pedagogical Topics by Student
Score
Groups}\label{relative-prevalence-of-pedagogical-topics-by-student-score-groups}

\pandocbounded{\includegraphics[keepaspectratio]{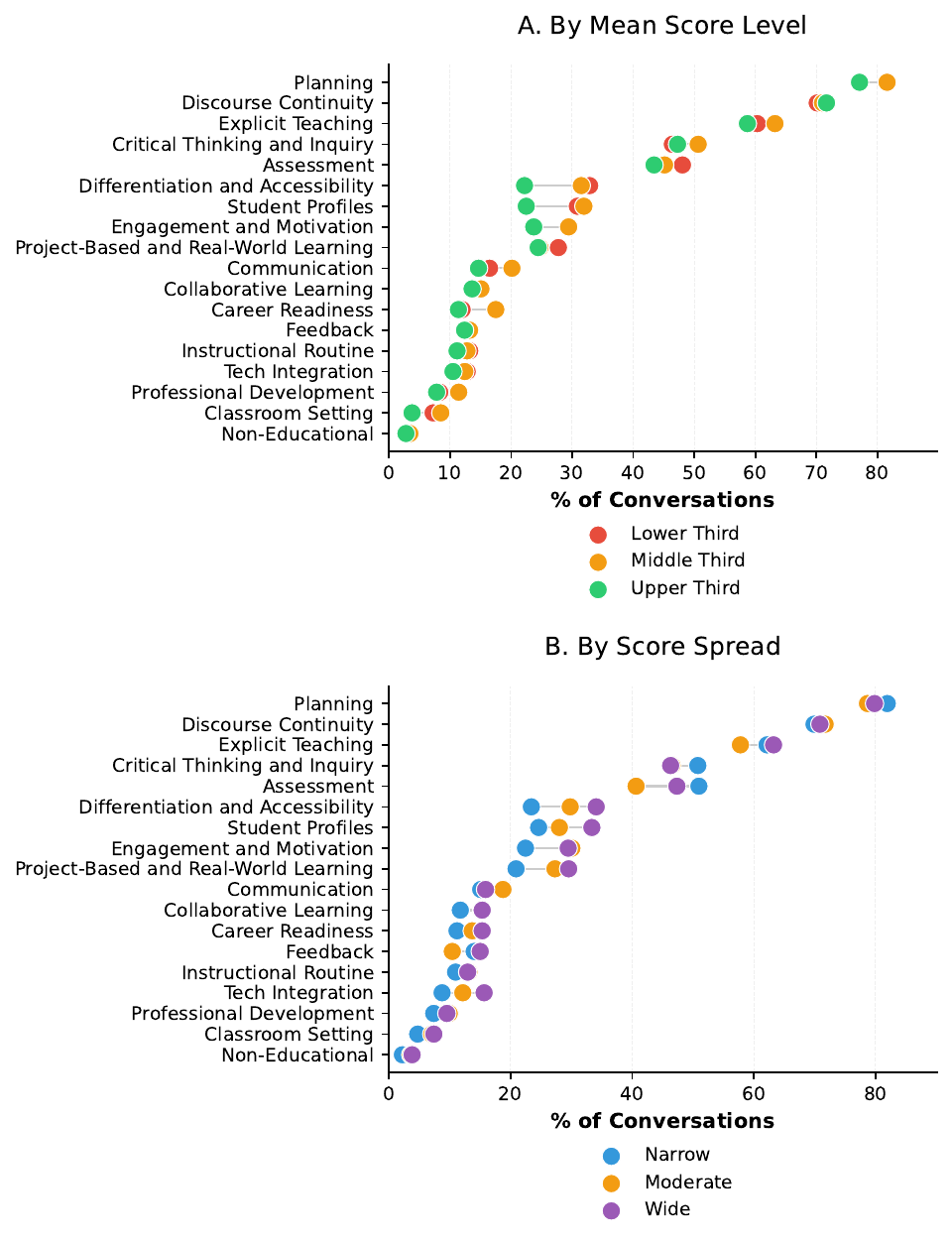}}

\newpage{}

\section{Student Interim Test Scores \& Teacher Usage
Correlations}\label{student-interim-test-scores-teacher-usage-correlations}

This section pools 12,005 students with matched beginning-of-year (BOY)
and middle-of-year (MOY) interim assessment scores across 2
districtsthat provided such data. BOY assessments were primarily
administered in September 2025 and MOY assessments were administered
primarily in December 2025 and January 2026. The exact days of
administration varied across grade and subjects. Assessment data came
from the STAR Math and Reading exams as well as the i-Ready Math and
Reading exams. The analysis examines whether students whose teachers
used Colleague AI more intensively showed different score trajectories,
controlling for prior scores, student characteristics, and accounting
for differences across assessment types and districts.

\begin{table}[H]
\centering
\footnotesize
\begin{tabular}{lcccc}
\hline\hline
 & \multicolumn{2}{c}{Math} & \multicolumn{2}{c}{Reading} \\
\cmidrule(lr){2-3} \cmidrule(lr){4-5}
 & Mean & SD & Mean & SD \\
\hline
\hline
SPED & 0.15 & 0.35 & 0.14 & 0.35 \\
ELL & 0.17 & 0.38 & 0.17 & 0.38 \\
Attendance Rate & 0.85 & 0.11 & 0.85 & 0.11 \\
AI Hours & 0.71 & 0.85 & 0.71 & 0.84 \\
AI Hours (Active Tchrs) & 2.97 & 4.18 & 2.91 & 3.91 \\
Num Messages & 15.73 & 17.41 & 15.80 & 17.26 \\
Num Msgs (Active Tchrs) & 66.48 & 85.62 & 65.43 & 81.12 \\
Num Linked Teachers & 3.86 & 3.11 & 3.93 & 3.15 \\
\hline
N Students & \multicolumn{2}{c}{11,259} & \multicolumn{2}{c}{11,292} \\
\quad Grade 3 & \multicolumn{2}{c}{1,905} & \multicolumn{2}{c}{1,879} \\
\quad Grade 4 & \multicolumn{2}{c}{1,963} & \multicolumn{2}{c}{1,941} \\
\quad Grade 5 & \multicolumn{2}{c}{1,836} & \multicolumn{2}{c}{1,791} \\
\quad Grade 6 & \multicolumn{2}{c}{1,934} & \multicolumn{2}{c}{1,945} \\
\quad Grade 7 & \multicolumn{2}{c}{1,869} & \multicolumn{2}{c}{1,904} \\
\quad Grade 8 & \multicolumn{2}{c}{1,752} & \multicolumn{2}{c}{1,832} \\
\hline\hline
\end{tabular}
\vspace{1ex}
\parbox{0.8\textwidth}{\scriptsize Statistics computed on the regression analysis sample (students with matched BOY and MOY scores). Student-level variables deduplicated by student within each subject.}
\end{table}

The first table presents summary statistics for the regression sample,
broken down by subject. Because the analysis pools multiple assessments
across districts, BOY and MOY scores are omitted from the summary;
scores are standardized (z-scored) within assessment, grade, and subject
before entering the regression. Sample sizes differ across subjects
because not all students have matched BOY and MOY scores in every
subject. The AI hours row reflects the average Colleague AI usage among
tested students' linked teachers; because many teachers used the
platform lightly or not at all during this early adoption period, the
average is modest. The active-teacher rows show the mean hours or
messages among teachers who logged at least one session.

\begin{table}[H]
\centering
\footnotesize
\begin{tabular*}{\textwidth}{@{\extracolsep{\fill}}lcccccccc}
\hline\hline
 & \multicolumn{4}{c}{Math} & \multicolumn{4}{c}{Reading} \\
\cmidrule(lr){2-9}
 & \multicolumn{2}{c}{AI Hrs} & \multicolumn{2}{c}{Num Msgs} & \multicolumn{2}{c}{AI Hrs} & \multicolumn{2}{c}{Num Msgs} \\
 & (1) & (2) & (3) & (4) & (5) & (6) & (7) & (8) \\
\hline
AI Hrs & -0.0004 & -0.0008 &  &  & 0.0094* & 0.0090* &  &  \\
 & (0.0054) & (0.0054) &  &  & (0.0050) & (0.0049) &  &  \\
Num Msgs &  &  & 0.0001 & 0.0001 &  &  & 0.0005** & 0.0005** \\
 &  &  & (0.0003) & (0.0003) &  &  & (0.0002) & (0.0002) \\
Prior & 0.8715*** & 0.8380*** & 0.8717*** & 0.8380*** & 0.8968*** & 0.8564*** & 0.8968*** & 0.8565*** \\
 & (0.0046) & (0.0053) & (0.0046) & (0.0053) & (0.0042) & (0.0050) & (0.0042) & (0.0050) \\
\hline
Stu ELL &  & -0.1462*** &  & -0.1461*** &  & -0.1358*** &  & -0.1358*** \\
 &  & (0.0133) &  & (0.0133) &  & (0.0122) &  & (0.0122) \\
Stu SPED &  & -0.0863*** &  & -0.0861*** &  & -0.1257*** &  & -0.1255*** \\
 &  & (0.0136) &  & (0.0136) &  & (0.0129) &  & (0.0129) \\
Attendance &  & 0.1202*** &  & 0.1215*** &  & 0.1871*** &  & 0.1880*** \\
 &  & (0.0443) &  & (0.0443) &  & (0.0399) &  & (0.0399) \\
\hline
R$^2$ & 0.762 & 0.765 & 0.762 & 0.765 & 0.804 & 0.808 & 0.804 & 0.808 \\
N & 11,259 & 11,259 & 11,259 & 11,259 & 11,292 & 11,292 & 11,292 & 11,292 \\
\hline\hline
\multicolumn{9}{l}{\scriptsize Standard errors in parentheses. * p $<$ 0.1, ** p $<$ 0.05, *** p $<$ 0.01. All models include assessment type and district fixed effects.} \\
\end{tabular*}
\end{table}

This table reports OLS regression results predicting middle-of-year
(MOY) standardized test scores. All models control for beginning-of-year
(BOY) scores, which are strongly predictive of MOY performance, and
include assessment type and district fixed effects. For each subject, we
estimate two specifications: a base model with only prior scores and the
treatment variable, and a controlled model that adds student-level
covariates for English Language Learner status, disability status, and
attendance rate. We report results for two measures of teacher AI use:
total hours on the platform and total messages sent, aiming to assess
robustness across different operationalizations of AI engagement.

In reading, teacher AI hours show a small, statistically significant
positive association with student MOY scores in both the base and
controlled specifications. This association is not present in
mathematics. The effect size is modest, on the order of 0.025 standard
deviations at the average teacher AI usage level of 3 hours, but is
notable given that this reflects only the first four months of a
voluntary adoption period during which most teachers were still learning
the platform. These are correlational findings and should be interpreted
with caution; teachers who chose to use Colleague AI more intensively
may differ from lighter users in unobserved ways. The consistency of the
reading result across specifications and treatment measures provides an
encouraging early signal that we will continue to investigate as the
study progresses through the full 2025-26 school year.

To effectively contextualize the 0.025 standard deviation effect size
observed in reading, it is helpful to view this early finding alongside
established benchmarks in educational research. Hattie (2008) notes that
the average effect size for educational interventions is approximately
0.40 standard deviations, a benchmark that typically represents a full
year of intensive, targeted instruction. Furthermore, Guskey and Yoon
(2009) emphasize that translating teacher learning and new practices
into measurable student achievement usually requires sustained
engagement, often identifying a threshold of 30 or more contact hours to
see significant effects. The current study's findings reflect only three
average hours of voluntary teacher engagement over a limited four-month
period. Additionally, rigorous evaluations of educational technology
frequently reveal negligible short-term impacts. For instance,
Campuzano, Dynarski, Agodini, and Rall (2009) found that several widely
adopted reading and mathematics software products produced effect sizes
that were not statistically different from zero. Given the historical
challenges of demonstrating rapid, measurable gains from new educational
technology integrations, the modest but statistically significant
positive association in reading provides an early baseline relative to
the low dosage of teacher platform usage. We will explore these
preliminary findings in greater detail, utilizing a more complete
dataset and rigorous analytical methods, in the forthcoming end-of-year
report.

\newpage{}

\section{References}\label{references}

Campuzano, L., Dynarski, M., Agodini, R., \& Rall, K. (2009).
Effectiveness of Reading and Mathematics Software Products: Findings
From Two Student Cohorts. NCEE 2009-4041. National Center for Education
Evaluation and Regional Assistance.

Esbenshade, L., Sarkar, S., Nucci, D., Edwards, A., Nielsen, S.,
Rosenberg, J. M., \ldots{} \& He, K. (2025). Emerging Patterns of GenAI
Use in K-12 Science and Mathematics Education. arXiv preprint
arXiv:2509.10747.

Guskey, T. R., \& Yoon, K. S. (2009). What works in professional
development?. Phi delta kappan, 90(7), 495-500.

Hattie, J. (2008). Visible learning: A synthesis of over 800
meta-analyses relating to achievement. routledge.

Liu, A., Esbenshade, L., Sarkar, S., Tian, V., Zhang, Z., He, K., \&
Sun, M. (2025). How K-12 Educators Use AI: LLM-Assisted Qualitative
Analysis at Scale. arXiv preprint arXiv:2507.17985.

NCES. (2024). Students with Disabilities. \emph{Condition of Education}.
U.S. Department of Education, Institute of Education Sciences.
\url{https://nces.ed.gov/programs/coe/indicator/cgg/students-with-disabilities}

OSPI. (2025). Transitional Bilingual Instruction Program: Report to the
Legislature, 2024--25.
\url{https://ospi.k12.wa.us/sites/default/files/2026-01/tbip-legislative-report-2024-25.pdf}

Pew Research Center. (2023). What federal education data shows about
students with disabilities in the U.S.
\url{https://www.pewresearch.org/short-reads/2023/07/24/what-federal-education-data-shows-about-students-with-disabilities-in-the-us/}

Reardon, S. F. (2011). The Widening Academic Achievement Gap Between the
Rich and the Poor: New Evidence and Possible Explanations. Whither
Opportunity?: Rising Inequality, Schools, and Children's Life Chances,
91.

Turgut, R., Sahin, S., \& Huerta, M. (2016). Teachers' perceptions of
effective school-wide programs and strategies for English language
learners. \emph{Learning Environments Research}, 19(2), 175--194.
\url{https://doi.org/10.1007/s10984-015-9173-6}

\section{Appendix: Report Data Sources}\label{sec-appendix-datasources}

The AmplifyLearn.AI research team would like to thank the districts that
provided mid-year administrative data. In follow up reports we expect to
include administrative data from a wider set of participating districts.
Because only one district was able to provide teacher administrative
data at this time, we have excluded that section of analysis from this
report.

District-provided administrative files vary in availability:

\begin{longtable}[]{@{}lr@{}}
\toprule\noalign{}
Data Type & Districts Providing \\
\midrule\noalign{}
\endhead
\bottomrule\noalign{}
\endlastfoot
Teacher Administrative Data & 1 of 12 \\
Student Demographics & 4 of 12 \\
Assessment Data & 3 of 12 \\
Student Attendance & 2 of 12 \\
\end{longtable}

\begin{itemize}
\tightlist
\item
  Teacher Administrative Data: this includes teachers' years of
  experience, certificate status, and whether or not they have earned a
  masters degree.
\item
  Student Demographics: this requested file includes students' grade
  level, and their binary multi-lingual and Special Education learner
  designation.
\item
  Assessment Data: this file includes information on standardized
  assessment scores including state summative tests like the SBA and
  interim assessments administered by the district.
\item
  Student Attendance: this file includes information on student
  attendance data.
\end{itemize}

Colleague AI platform usage data was provided by Colleague AI. All data
is stored securely and id variables that link platform usage records to
district provided data are de-identified before being accessed by the
AmplifyLearn.AI research team.

\newpage{}

\section{Appendix: Explanation of Feature
Categories}\label{sec-appendix-features}

\subsection{Lesson Planning}\label{lesson-planning}

\begin{itemize}
\tightlist
\item
  \textbf{Claire AI}: The primary teacher-facing AI assistant that
  powers Colleague AI's content generation tools, including Brainstorm
  Ideas conversations, lesson plan creation, interactive editing, and
  auto-generating start messages for Teaching Aides.
\item
  \textbf{Charlie AI}: This is a secondary teacher-facing AI assistant
  that is used to test experimental changes to the AI chat interface
  before they are implemented in the main Claire AI.
\item
  \textbf{Generate Lesson}: Creates comprehensive, standards-aligned
  lesson plans based on teacher-provided learning objectives, subject,
  grade level, differentiation needs, and state standards.
\item
  \textbf{Lesson Plan Conversation}: This feature encompasses AI
  conversations that teachers have using a dedicated sidebar when
  viewing a lesson plan document in the My Documents area of Colleague
  AI.
\item
  \textbf{Quality Measures}: Evaluates lesson plan quality using
  research-validated rubrics (developed via Delphi method with
  practitioners), allowing teachers to compare their professional
  judgment with AI-generated reviews.
\item
  \textbf{In-text AI Command}: When viewing documents in My Documents,
  teachers have the ability to highlight specific lines of text and
  engage in an AI conversation for a quick edit on that specific portion
  of the text.
\end{itemize}

\subsection{Lesson Delivery}\label{lesson-delivery}

\begin{itemize}
\tightlist
\item
  \textbf{Generate Image}: Produces custom instructional images in
  various styles (realistic, cartoon, diagram) from text descriptions,
  with prompt history and regeneration support.
\item
  \textbf{Generate Interactive}: Builds engaging online student
  activities such as quizzes, worksheets, drag-and-drop exercises, and
  formative assessments that can be shared via link and report on
  student responses.
\item
  \textbf{Generate Slides}: Creates presentation slide decks from topic
  and learning objectives, with support for both Colleague AI's template
  library and custom teacher-uploaded templates.
\item
  \textbf{Generate Podcast}: Produces audio learning content from
  teacher-specified topics, with controls for target audience, desired
  length, and tone (conversational, formal, storytelling).
\item
  \textbf{Generate Diagram}: This feature creates editable diagrams for
  teachers.
\item
  \textbf{Lesson Plan Simulation}: A lesson plan enhancement tool that
  lets teachers preview and simulate how a generated lesson might unfold
  in the classroom before teaching it.
\end{itemize}

\subsection{Student Interaction \&
Feedback}\label{student-interaction-feedback}

\begin{itemize}
\tightlist
\item
  \textbf{Generate Rubric}: Creates assessment rubrics aligned to
  learning standards with customizable criteria dimensions and
  performance levels (e.g., Exceeds, Meets, Approaching, Beginning).
\item
  \textbf{AI Grading}: An AI-powered grading assistant that scores
  student submissions against teacher-provided rubrics and answer keys,
  generating detailed rubric-based feedback with export options.
\item
  \textbf{AI Tutor}: \emph{Student Facing} An open-ended,
  student-initiated AI conversation feature within a classroom that
  allows students to ask questions on any topic, with conversations
  visible to the teacher.
\item
  \textbf{Teaching Aide}: \emph{Student Facing} A teacher-designed,
  topic-specific AI assistant deployed to students within a classroom,
  where teachers control the prompt, uploaded materials, and monitor
  conversations in real time.
\end{itemize}

\section{Appendix: Explanation of Pedagogical
Topics}\label{sec-appendix-topics}

Definitions reprinted with permission from Liu, A., Esbenshade, L.,
Sarkar, S., Tian, V., Zhang, Z., He, K., \& Sun, M. (2025). How K-12
Educators Use AI: LLM-Assisted Qualitative Analysis at Scale. arXiv
preprint arXiv:2507.17985.

\begin{itemize}
\item
  \textbf{Assessment.} Educators requested assistance in generating
  formative assessments, summative tasks, and grading rubrics, typically
  aligned with specific instructional units or learning objectives.
\item
  \textbf{Career Readiness.} A subset of prompts focused on supporting
  students' exploration of career pathways, college readiness, or
  workplace skills, integrating postsecondary preparation into lesson
  and activity planning. These subthemes also align with the Curriculum
  and Content Focus domain, as such conversations often involved
  generating informational materials and instructional resources.
\item
  \textbf{Classroom Settings.} Prompts referenced specific instructional
  contexts, including homeschooling, low-tech environments, afterschool
  programs, and classrooms requiring behavioral interventions. These
  contextual codes suggest that educators use generative AI to adapt
  instruction in response to structural constraints as well as
  individual learner needs.
\item
  \textbf{Communication.} Requests related to drafting communications
  with families, colleagues, or administrators constituted a substantial
  portion of educator usage. These interactions demonstrate that AI was
  used not only for student-facing instructional tasks but also for
  administrative and professional communication.
\item
  \textbf{Critical Thinking and Inquiry.} Many prompts asked AI to
  support strategies that encouraged deep questioning, research and
  source analysis, and historical, civic, or economic reasoning.
  Educators frequently sought guidance on structuring inquiry-based
  learning experiences and fostering higher-order thinking.
\item
  \textbf{Differentiation and Accessibility.} Educators commonly sought
  support for differentiated instructional strategies tailored to
  English Language Learners (ELLs), students with Individualized
  Education Programs (IEPs), and mixed-ability classrooms. Prompts
  frequently included requests for tiered scaffolding, visual
  representations, and multilingual adaptations, particularly in core
  content areas.
\item
  \textbf{Discourse Continuity.} Discourse Continuity in educator-AI
  conversation refers to the extent to which the interaction builds
  coherently across turns, with the teacher treating prior AI responses
  as part of an ongoing exchange rather than as isolated outputs. It
  captures whether the teacher follows up on earlier ideas, asks the AI
  to continue or refine prior work, rejects unsatisfactory output, or
  requests changes in content or format while maintaining the same
  underlying conversational thread. In this sense, discourse continuity
  reflects sustained engagement with the evolving response and the
  teacher's active orchestration of the interaction.
\item
  \textbf{Explicit Teaching.} Educators regularly prompted the AI to
  model or explain core concepts in STEM and literacy, including
  explaining mathematical concepts, modeling problem-solving processes,
  and supporting English language arts skill development. These requests
  reflect efforts to strengthen direct instruction in foundational
  content areas.
\item
  \textbf{Feedback.} Educators used AI to draft feedback, interpret
  patterns in student performance, and suggest next instructional steps.
  AI often played an argumentative and elaborative role in tasks such as
  generating feedback for students or supporting data-driven progress
  monitoring.
\item
  \textbf{Instructional Routines and Engagement.} Some prompts
  emphasized learning progression and routine adjustments, while others
  focused on actionable strategies to increase student motivation. These
  requests were particularly common in early-grade contexts and
  intervention-oriented settings, highlighting the role of AI in
  supporting classroom pacing and engagement.
\item
  \textbf{Planning.} Planning-related conversations appeared prominently
  in educator requests. Educators frequently engaged with generative AI
  to prototype full lesson sequences, align instruction with learning
  standards, and design or adjust in-class activities. These
  planning-oriented requests were often combined with subject-specific
  scaffolding and instructional considerations, highlighting the close
  interdependence between content planning and adaptive pedagogy. The
  co-occurrence of planning with instructional strategies suggests that
  educators use AI not only to organize curricular materials but also to
  refine how those materials are enacted in classroom instruction.
\item
  \textbf{Professional Development.} A smaller subset of prompts focused
  on reflective practice and the preparation of professional learning or
  workshop materials, indicating AI use in supporting educators' own
  learning and growth.
\item
  \textbf{Project-Based and Real-World Learning.} Educators leveraged AI
  to generate tasks aligned with real-world engagement, hands-on
  projects, experimental design, and technical skill development. These
  practices were often connected to authentic disciplinary applications,
  such as engineering design challenges and science laboratory
  investigations, reflecting an emphasis on applied learning.
\item
  \textbf{Student Profiles.} Educators frequently described target
  learners as below grade level, gifted, English Language Learners
  (ELLs), students receiving special education services, or students
  with social--emotional support needs. These descriptors were often
  paired with requests for differentiated instructional strategies,
  highlighting a common linkage between pedagogical approach and learner
  profile.
\item
  \textbf{Technology Integration.} Educators also used AI to explore
  questions related to the integration of technology and multimedia
  tools into instruction. In addition to inquiries about using AI
  itself, prompts included requests for incorporating digital resources
  and multimedia to support teaching and learning. Notably, more
  systematic and infrastructure-oriented technology implementation
  conversations appeared in interactions initiated by administrators,
  reflecting broader organizational considerations around instructional
  technology adoption.
\end{itemize}

\end{document}